\documentclass[10pt]{article} 
\usepackage[accepted]{rlc}
\usepackage{booktabs}
\usepackage{float}
\usepackage{amssymb}            
\usepackage{mathtools}          
\usepackage{mathrsfs}           
\mathtoolsset{showonlyrefs}     
\usepackage{graphicx}           
\usepackage{subfigure}
\graphicspath{{./figures/}}
\usepackage{subcaption}         
\usepackage[space]{grffile}     
\usepackage{url}                

\title{
    \centering
    \begin{minipage}{0.85\textwidth} 
        \centering
        \textbf{SkillPager: Query-Adaptive Intra-Skill Navigation via Semantic Node Retrieval
        }
    \end{minipage}
}

\author{
    \hspace*{-2em}
    \begin{minipage}{\linewidth}
        \centering
        \textbf{
            Zicai Cui\textsuperscript{1}, 
            Zihan Guo\textsuperscript{2,3}, 
            Weiwen Liu\textsuperscript{1}, 
            Weinan Zhang\textsuperscript{1,2} 
        }
        \\[1em] 
        \normalfont
        \textsuperscript{1}Shanghai Jiao Tong University \quad
        \textsuperscript{2}Shanghai Innovation Institute 
        \quad
        \textsuperscript{3}Sun Yat-sen University 
        \\[0.5em] 
        \texttt{zicaicui@sjtu.edu.cn} \quad 
    \end{minipage}
}

\begin{document}

\maketitle

\begin{abstract}
Skill-based LLM agents increasingly rely on long procedural documents, but full-document prompting wastes tokens and dilutes information critical to execution. We study this setting as \emph{intra-skill retrieval}, where the goal is to select a minimal, execution-sufficient context from a known skill document given a query. We present \textsc{SkillPager}, a two-stage framework that parses each Markdown skill into typed semantic nodes offline and leverages Maximal Marginal Relevance (MMR) to perform global, query-conditioned node selection online. On a benchmark of 395 skills and 1,975 queries, \textsc{SkillPager} achieves 78.89\% LLM-judged context sufficiency, compared to 82.23\% for the exhaustive full-document baseline, while reducing prompt tokens by 47.04\%. A granularity ablation shows that applying the same retrieval algorithm to raw fixed-length chunks reaches a comparable 81.77\% sufficiency but increases token cost by 28.81\%, demonstrating that efficiency gains are driven by typed semantic granularity rather than the retrieval algorithm alone. Among graph-based baselines, \textsc{SkillPager} outperforms the strongest baseline by a margin of 12.16\%. Further ablations show that supporting content is most effective when retained in the candidate pool and selected adaptively rather than removed by static heuristics. These results identify typed intra-document retrieval as a distinct access problem for skill-based agents.

\end{abstract}

\section{Introduction}
\label{sec:intro}

Skills are increasingly used as the interface through which LLM agents access reusable procedural knowledge~\citep{shen2023hugginggpt,qin2024toolllm,zhao2024expel}. In practice, however, skills are stored as long Markdown documents that mix executable steps with examples, parameter specifications, tool-specific guidance, and conceptual background. Consequently, injecting the full document at every invocation becomes prohibitively expensive and degrades execution performance, as critical instructions are easily diluted by such peripheral content~\citep{liu2024lost,jiang2023llmlingua}. The resulting problem is not standard corpus retrieval, but intra-skill retrieval: given a single known skill document and an execution query, construct a compact context that preserves the information necessary for task execution.

\begin{figure}[t]
  \centering
  \includegraphics[width=\linewidth]{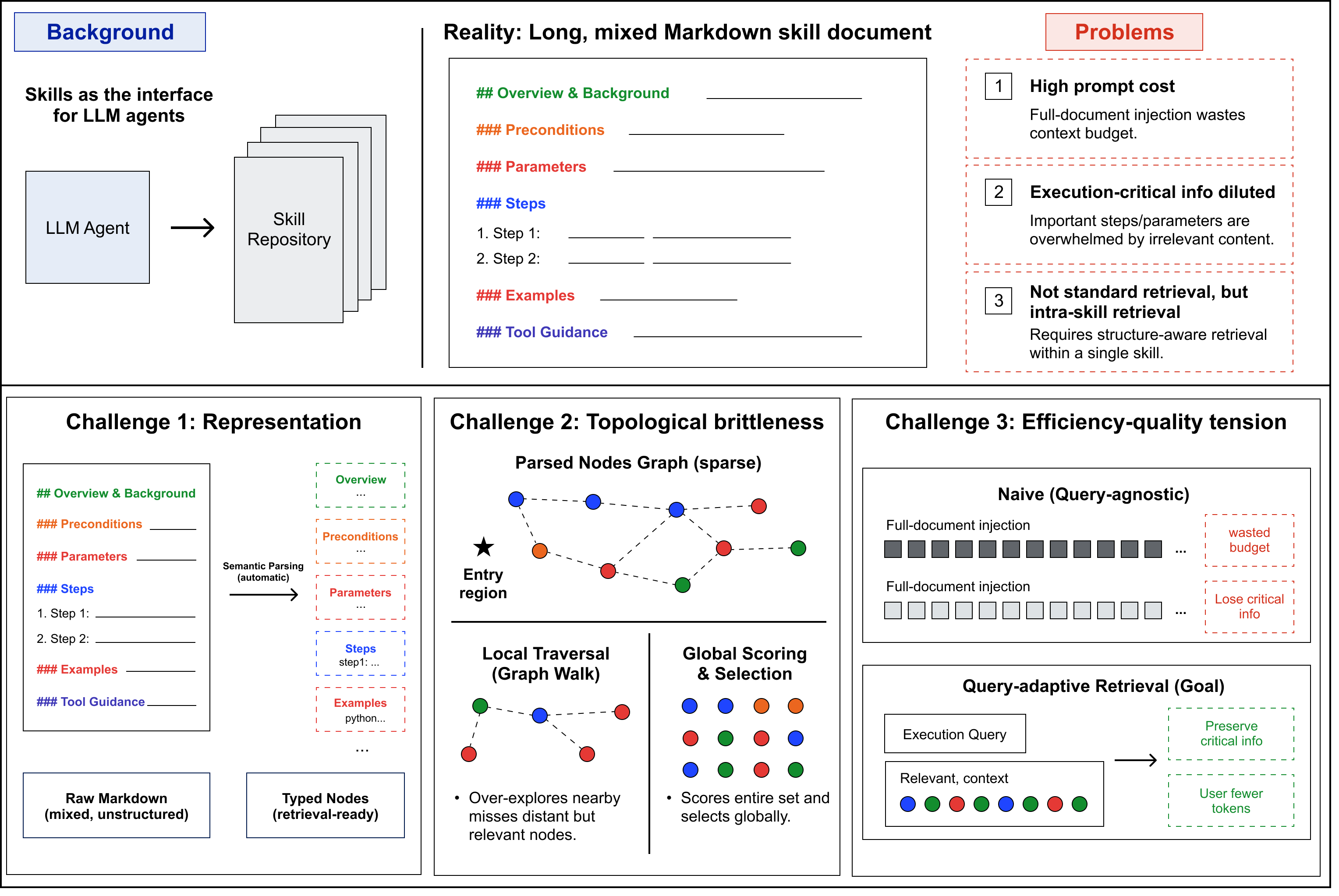}
  \caption{Background and core challenges of intra-skill retrieval. Given a query and a known skill document, the system must retrieve a compact, execution-sufficient context from a long, mixed Markdown skill rather than inject the full document. This requires (1) converting raw skill text into retrieval-ready semantic units, (2) avoiding brittle local traversal over sparse parsed graphs, and (3) balancing context sufficiency against prompt efficiency.}
  \label{fig:challenges}
\end{figure}

Intra-skill retrieval relies on two core components. The first is a retrieval-ready representation, where a raw skill document must be converted into semantically addressable units before any selection can occur. The second is a query-adaptive retrieval strategy designed to select these units globally and efficiently, maximizing context sufficiency while minimizing token overhead. As illustrated in Figure~\ref{fig:challenges}, realizing these components introduces three challenges, which we detail below.

\paragraph{Challenge 1: Representation.}
Raw skill documents are typically authored in free-form Markdown, lacking retrieval-ready structure. Within skill documents, crucial execution semantics (such as operational steps and parameter constraints) are deeply interleaved with non-executable content like commentary and explanatory text. Efficient intra-skill retrieval therefore requires structured semantic parsing that granularly deconstructs the document into distinct, typed nodes without relying on laborious manual annotation (\S\ref{sec:d1}).

\paragraph{Challenge 2: Topological brittleness in retrieval.}
An intuitive baseline approach to managing these parsed units is to construct and traverse a localized dependency graph. However, we observe that such document-derived graphs are inherently sparse and unreliable: execution-critical nodes are frequently disconnected or topologically distant from the document's entry point. Sequential or graph-based local traversal thus suffers from severe \emph{entry bias}, over-exploring localized neighborhoods while missing distant yet vital context. Because the core bottleneck stems from topology quality rather than traversal depth, intra-skill retrieval necessitates scoring the \emph{entire} ungrounded node set globally (\S\ref{sec:d2}).

\paragraph{Challenge 3: Efficiency--quality tension in retrieval.}
Static strategies such as full-document injection and fixed-window truncation ignore the query and waste budget on irrelevant content. A good intra-skill retriever must adapt its selection to the query, preserving information critical to task completion while substantially reducing token consumption (\S\ref{sec:results}).

The key observation is that skill documents are procedurally ordered but topologically sparse. Consequently, graph topology carries limited retrieval signal, making global selection far more effective than local traversal. Once a skill is decomposed into typed semantic nodes, intra-skill retrieval can be cast as a global MMR problem~\citep{carbonell1998use}: sparse graphs make traversal brittle, while typed nodes make relevance-diversity optimization meaningful. Prior work either compresses offline into a fixed artifact (SkillReducer~\citep{gao2026skillreducer}) or requires manually authored edges (ObjectGraph \citep{dubey2026objectgraph}), neither of which provides query-adaptive global selection over typed, unannotated nodes.

We instantiate this idea in \textsc{SkillPager}, a two-stage framework. In Stage~1 (offline), a semantic parsing pipeline converts raw Markdown into typed nodes spanning six roles: \textit{step}, \textit{example}, \textit{param}, \textit{precondition}, \textit{error\_handling}, and \textit{concept}. In Stage~2 (online), global MMR selection operates over the full node set with a dynamic budget scaled adaptively based on the total node count. 

Our contributions are threefold:
\begin{itemize}
  \item We present, to our knowledge, the first formalization of intra-skill retrieval as a standalone problem for skill-based agents, decoupling it from cross-skill routing and general corpus retrieval.
  
  \item We propose \textsc{SkillPager}, a two-stage framework that pairs offline typed semantic parsing with online query-adaptive global MMR selection. This topology-independent design enables the system to reliably extract execution-critical content without relying on fragile local graph traversal.

  \item We build \textsc{IntraSkill-Bench}, a controlled evaluation dataset featuring 395 skills and 1,975 queries with a human-validated LLM-judge protocol. Empirical results demonstrate that SkillPager achieves 78.89\% context sufficiency while reducing prompt tokens by 47.04\%, approaching the full-document reference at nearly half the token cost and significantly outperforming graph-based baselines.
\end{itemize}

The remainder of the paper is organized as follows.~\S\ref{sec:relate} reviews relevant work in areas including context construction and compression, agent skill management, and diversity-aware retrieval.~\S\ref{sec:method} presents the semantic parsing pipeline and the global MMR retrieval procedure.~\S\ref{sec:experiments-setup} describes the experimental setup,~\S\ref{sec:results} reports the empirical results and analysis, and~\S\ref{sec:conclusion} concludes.

\section{Related Work}
\label{sec:relate}

\subsection{Retrieval-Augmented Generation and Context Compression}

Retrieval-augmented generation (RAG) grounds large language models in external knowledge by retrieving evidence from a corpus and conditioning generation on the retrieved content~\citep{lewis2020retrieval}. A central foundation is dense passage retrieval, which encodes queries and passages into a shared embedding space and matches them by similarity~\citep{karpukhin2020dense}. This formulation made retrieval practical at scale and shaped many later systems that combine retrieval with generation.

As retrieval systems increase the evidence passed to the model, long prompts raise inference cost and can degrade performance when salient evidence is diluted by peripheral content. This has motivated a line of work on prompt compression. LLMLingua introduces a coarse-to-fine framework with budget control and token-level iterative pruning~\citep{jiang2023llmlingua}, while LongLLMLingua extends this approach to long-context settings~\citep{jiang2024longllmlingua}. xRAG pushes compression further by fusing dense retrieval embeddings directly into the language model representation space~\citep{cheng2024xrag}. These approaches operate at the level of prompt text or document representations and remain document-level in form.

SkillReducer studies context compression in the setting of agent skills and performs offline compression of skill descriptions and bodies~\citep{gao2026skillreducer}. Its compressed artifact, however, is fixed once constructed and does not adapt to individual queries at inference time. \textsc{SkillPager} addresses the same efficiency pressure from a different angle: instead of compressing a skill document into a single shared artifact, it performs query-adaptive selection over semantic units within a known skill document.

\subsection{Agent Skill Management}

Agent systems increasingly rely on skills, i.e., structured procedural documents that encode reusable capabilities, tool usage patterns, and execution guidance. Early work on tool-augmented language models, such as Toolformer, showed that models can learn when and how to invoke external APIs~\citep{schick2023toolformer}. That line of work, however, typically assumes a small, fixed tool set rather than a large and evolving repository of reusable skill documents. As skill libraries grow, agent systems face both a routing problem (which skill to use) and an efficiency problem (how skill content should be represented and accessed).

Recent work has begun to study this space through structured representations. ObjectGraph introduces a typed graph-based file format for agent documents and performs graph traversal at inference time~\citep{dubey2026objectgraph}. Graph of Skills constructs a skill dependency graph offline and retrieves bounded skill bundles through graph-based propagation at inference time~\citep{li2026graph}. Both results highlight the value of structure-aware access for large skill libraries. \textsc{SkillPager} is related to these efforts but targets a different setting: rather than retrieving among skills or traversing a manually authored graph structure, it focuses on query-adaptive selection within a single known skill document.

\subsection{Selective Context Construction}

Selective context construction studies how to assemble a compact, task-relevant context from a larger candidate pool. Recent work has shown that flat passage retrieval is not the only viable retrieval substrate. RAPTOR imposes a recursive summary tree and retrieves across abstraction levels rather than from a flat chunk list~\citep{sarthi2024raptor}. Dense X Retrieval studies retrieval granularity directly and shows that finer units such as propositions can outperform passage-level indexing~\citep{chen2024dense}. These methods are structurally closer to \textsc{SkillPager} than token-pruning methods because they change the retrieval unit itself. \textsc{SkillPager}'s units, however, differ from summary-tree nodes or decontextualized propositions in a key respect: each node preserves the original text boundary of its source fragment and carries an explicit procedural type label, making the selected units directly consumable as execution context rather than retrieval surrogates.

Graph-structured methods extend this line further by building explicit relational structure before retrieval. GraphRAG constructs an entity graph and community summaries to support query-focused summarization over a corpus~\citep{edge2024local}. HippoRAG organizes retrieved knowledge as graph-structured long-term memory, and ArchRAG performs hierarchical retrieval over attributed graph communities~\citep{gutierrez2024hipporag,Wang_2026}. These results support the broader claim that imposed structure can improve selective context construction. \textsc{SkillPager} differs in both scope and objective. It operates within a single known document rather than across entities or communities, and its goal is not corpus-level sensemaking or evidence aggregation, but assembling a procedurally sufficient local context for the current execution query. Its retrieval units are typed semantic nodes that preserve original text boundaries and encode procedural roles such as step, parameter, and error handling.

Token-level compression methods such as LLMLingua and LongLLMLingua operate at a different granularity~\citep{jiang2023llmlingua,jiang2024longllmlingua}. They remove tokens or phrases while treating the output as a variable-length text sequence. \textsc{SkillPager} selects intact nodes, preserving structural units that are operationally meaningful. For procedural execution, a complete step or parameter block is often more useful than a shorter but structurally incomplete fragment. These methods compress an existing prompt rather than select from an unprocessed document. Applying them here would require either injecting the full document first (defeating the goal of reducing cost) or pre-compressing each skill offline (reducing to SkillReducer's fixed-artifact approach~\citep{gao2026skillreducer}). Neither path yields a clean comparison under intra-skill retrieval, so we leave this direction for future work.

The main distinction is therefore not only that \textsc{SkillPager} works within a single document, but that its candidate pool is typed. Anonymous chunk-based methods treat all candidates as semantically equivalent spans distinguished only by content similarity. \textsc{SkillPager}'s candidates carry explicit semantic roles, such as step, example, parameter, precondition, error handling, and concept. This typed structure shapes both the coverage objective and the diversity criterion. The selection problem is not only which units are relevant, but whether the selected node set covers the procedural roles required by the query.

\subsection{Diversity-Aware Retrieval}

Diversity-aware retrieval traces back to maximal marginal relevance (MMR), which formalizes the trade-off between selecting highly relevant items and avoiding redundancy among already selected results \citep{carbonell1998use}. MMR ranks candidates by their marginal value with respect to the current selected set, encouraging the retrieved context to cover multiple aspects of an information need. This principle has been used in document retrieval, summarization, and multi-evidence selection, especially in settings where redundant evidence wastes limited context budget~\citep{Yu_2025}.

Recent RAG work has brought this idea more directly into generation systems. DF-RAG extends MMR to retrieval-augmented generation and dynamically balances relevance and diversity for each query at test time \citep{khan-etal-2026-df}. Its setting is cross-document question answering, where selection is performed over retrieved documents or chunks from multiple sources. These results suggest that diversity control remains useful even when dense retrieval already captures semantic relevance.

\textsc{SkillPager} builds on the same retrieval principle but applies it in a different setting. Rather than selecting among external documents or chunks, \textsc{SkillPager} performs intra-document selection within a single known skill document. Its retrieval units are typed semantic nodes rather than anonymous passages, and its objective is to assemble a compact procedural context for the current query. \textsc{SkillPager} can therefore be viewed as a diversity-aware retrieval method for structured intra-skill navigation rather than for cross-document evidence aggregation.

\section{Method}
\label{sec:method}

\textsc{SkillPager} follows the problem decomposition introduced in~\S\ref{sec:intro}. Stage~1 addresses Challenge~1 by converting raw Markdown skill documents into typed semantic nodes, establishing a retrieval-ready representation. Stage~2 subsequently tackles both Challenges~2 and~3 by performing global MMR selection over the full node set, maximizing context sufficiency for task completion while minimizing token consumption. The empirical validity of this joint design is thoroughly evaluated in~\S\ref{sec:results}.

\paragraph{Notation and problem formulation.}
Let $D$ denote a skill document and $q$ an execution query. We write $\mathcal{V}(D)$ for the set of semantic nodes parsed from $D$, and $k$ for the maximum number of nodes that may be selected. A selected subset is written as $S \subseteq \mathcal{V}(D)$, and $S^*$ denotes an optimal subset. We refer to the offline parsing stage as Stage~1 and to the online retrieval stage as Stage~2.

Given $D$ and $q$, \textsc{SkillPager} aims to select a compact subset $S \subseteq \mathcal{V}(D)$ that maximizes query-conditioned context sufficiency under a budget constraint. We formalize this objective as
\begin{equation}
  S^* = \arg\max_{\substack{S \subseteq \mathcal{V}(D) \\ |S| \leq k}} \Psi(q, S),
  \label{eq:objective}
\end{equation}
where $\Psi(q, S)$ denotes the latent sufficiency of context $S$ for query $q$.
Since $\Psi(q, S)$ depends on downstream execution and is not directly computable from the retrieval representation alone, we approximate it via a tractable surrogate.
The intuition is that a context is sufficient for a query when it (i)~contains information relevant to the query and (ii)~covers the diverse procedural aspects required for execution without collapsing into redundant content. 
The MMR naturally captures both desiderata, where the relevance term rewards query--node alignment, while the diversity term penalizes redundancy among already selected nodes. 
We therefore treat the accumulated MMR score under budget $k$ as a surrogate for $\Psi(q, S)$ and evaluate the resulting retrieval quality empirically.

\subsection{Stage~1: Execution Semantic Parsing}
\label{sec:d1}

\begin{figure}[t]
  \centering
  \includegraphics[width=\linewidth]{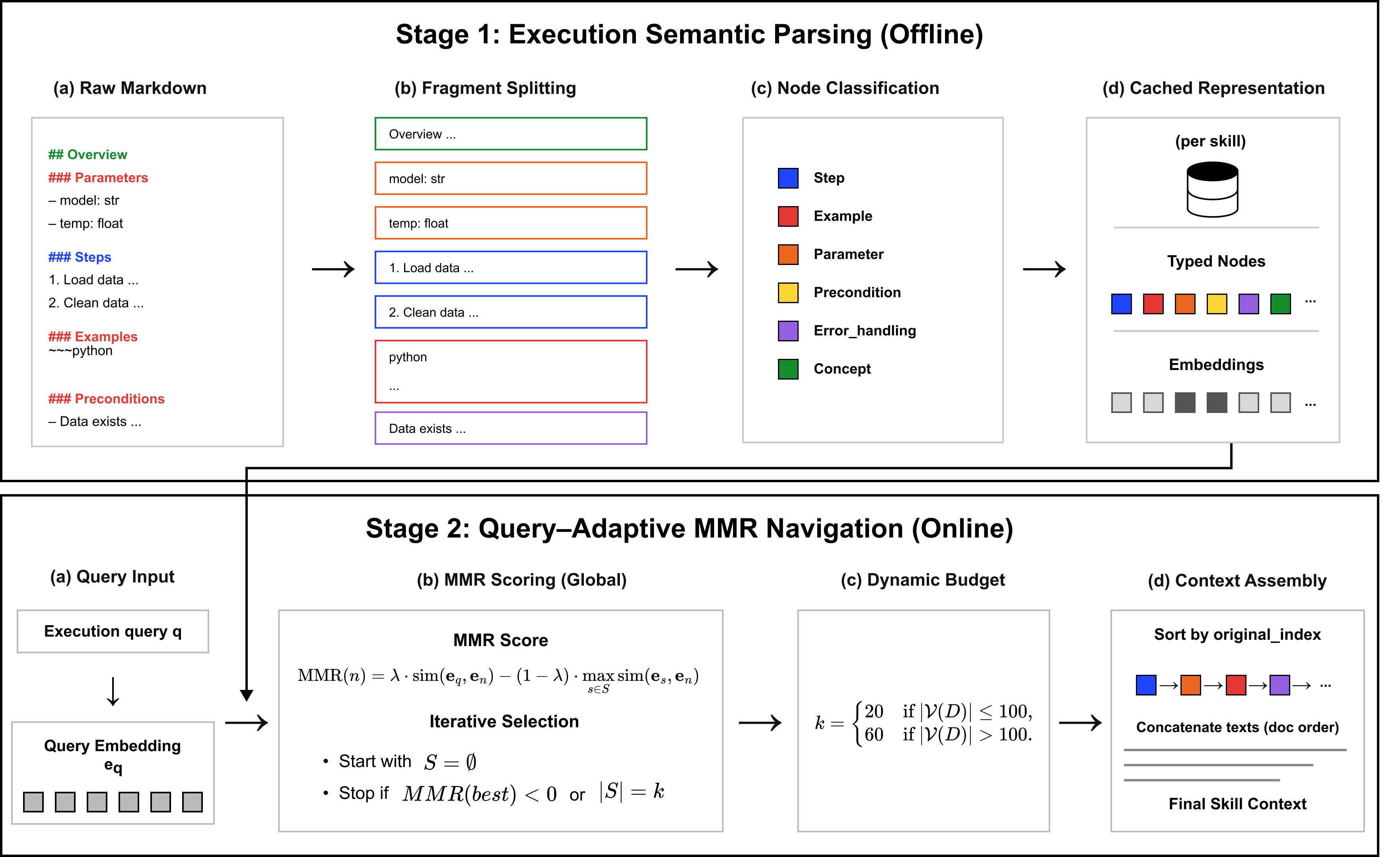}
  \caption{Overview of the \textsc{SkillPager} framework. Stage~1 performs offline execution semantic parsing by converting raw Markdown skill documents into retrieval-ready typed semantic nodes with cached embeddings. Stage~2 performs online query-adaptive MMR navigation over the cached node set to construct compact execution contexts through global relevance–diversity retrieval under a dynamic budget.}
  \label{fig:pipeline}
\end{figure}

Stage~1 converts a raw Markdown skill document into a typed set of semantic nodes that serves as the retrieval-ready representation used by \textsc{SkillPager} (Figure~\ref{fig:pipeline}). The objective is to expose semantically meaningful retrieval units while preserving the procedural structure of the source document.

\paragraph{Fragment Splitting.}
We first split the document into fragments at sentence boundaries, with rule-based merging to preserve local semantic coherence. Code blocks are kept intact. Sentences ending with a colon or a conditional opener, such as \textit{if}, \textit{when}, or \textit{otherwise}, are merged with the following content because they function as local headers. Connective fragments, such as \textit{or}, \textit{and}, \textit{e.g.}, and \textit{i.e.}, are merged into the preceding sentence. List items under the same header are grouped and annotated with list context.

\paragraph{Node Classification.}
A language model assigns each fragment to one of six semantic types:
\begin{itemize}
  \item \textbf{step}: an executable action or instruction that directly contributes to task completion;
  \item \textbf{example}: a usage demonstration, code snippet, or input/output pair;
  \item \textbf{param}: a configurable parameter, option, or argument specification;
  \item \textbf{precondition}: a requirement or constraint that must hold before execution;
  \item \textbf{error\_handling}: failure-mode guidance or a recovery procedure;
  \item \textbf{concept}: a definitional or explanatory passage that provides background semantics without itself constituting an executable action.
\end{itemize}

\noindent The first five types are treated as executable nodes. Concept nodes are non-executable, but remain in the retrieval candidate pool so that explanatory content can be selected when it helps ground the query. We evaluate this design in~\S\ref{sec:concept-ablation}.

\paragraph{Edge Inference.}
Stage~1 also infers four types of directed edges between nodes:
\begin{itemize}
  \item \textbf{sequence}: links between adjacent nodes in document order;
  \item \textbf{semantic}: links between node pairs whose embedding cosine similarity exceeds $\tau = 0.85$;
  \item \textbf{reference}: links from a node to another node whose identifier or key term it mentions, detected by string matching against node headings;
  \item \textbf{condition}: links from a precondition node to the step it constrains, identified from dependency cues in text.
\end{itemize}

\noindent These edges define a parsed node graph that is stored in the cache for graph-based baseline comparison and visualization. These topological dependencies are entirely bypassed during \textsc{SkillPager}'s online retrieval stage, which operates strictly over the unstructured node pool. We return to this design choice in~\S\ref{sec:design-rationale}.

\paragraph{Caching.}
The parsed node set, edge set, and node embeddings are computed once per skill document and cached persistently. On the 395-skill benchmark, building the Stage~1 cache requires 9.83 hours of wall-clock time with four parallel workers, yielding a median offline preprocessing time of approximately 3.2 minutes per skill. Once cached, query-time retrieval requires only a single query embedding and the MMR scoring loop, with no repeated LLM calls over the skill body. The resulting cost profile favors relatively stable repositories, where skills are parsed once and reused across many queries.

\subsection{Stage~2: Query-Adaptive MMR Navigation}
\label{sec:d2}

Stage~2 performs query-conditioned retrieval over the typed node set produced by Stage~1 (Figure~\ref{fig:pipeline}). Given a query $q$ and the cached node set $\mathcal{V}(D)$ with pre-computed embeddings, it selects up to $k$ nodes using Maximal Marginal Relevance. The candidate pool contains all node types and retrieval operates over this flat node set.

\paragraph{MMR Scoring.}
Let $\mathbf{e}_q$, $\mathbf{e}_n$, and $\mathbf{e}_s$ denote the unit-normalized embedding vectors of the query, the candidate node $n$, and an already selected node $s \in S$, respectively. At each iteration, we select the node that maximizes
\begin{equation}
  \mathrm{MMR}(n) = \lambda \cdot \mathrm{sim}(\mathbf{e}_q, \mathbf{e}_n)
  - (1-\lambda) \cdot \max_{s \in S} \mathrm{sim}(\mathbf{e}_s, \mathbf{e}_n),
  \label{eq:mmr}
\end{equation}
where $S$ is the set of already selected nodes. We use $\mathrm{sim}(\cdot,\cdot)$ to denote cosine similarity and $\lambda$ controls the relevance--diversity trade-off. For the first iteration, where $S=\emptyset$, the redundancy term is defined as zero. We set $\lambda=0.7$ based on the sensitivity analysis in~\S\ref{sec:lambda-sweep}.

\paragraph{Early Stopping.}
The MMR loop terminates when the best remaining candidate has negative marginal value,
\begin{equation}
\mathrm{MMR}(n^*) < 0.
\end{equation}
Under this condition, adding another node would increase redundancy more than query-relevant coverage under the current objective. In practice, early stopping is most visible at low $\lambda$ values, whereas at $\lambda=0.7$ the selected set typically reaches the budget limit.

\paragraph{Dynamic Budget.}
The maximum number of selected nodes adapts to the size of the skill:
\begin{equation}
  k =
  \begin{cases}
    20 & \text{if } |\mathcal{V}(D)| \le 100, \\
    60 & \text{if } |\mathcal{V}(D)| > 100.
  \end{cases}
  \label{eq:dynamic-k}
\end{equation}
This two-tier budget reflects the empirical observation that larger skills require more retrieved content to maintain adequate coverage, while smaller skills benefit from tighter budgets. We determine these operating points from the $k$-sweep analysis in~\S\ref{sec:k-sweep}.

\paragraph{Context Assembly.}
Following the iterative selection, the nodes within the optimal subset $S^*$ are re-sorted according to their original sequential order in the source document $D$. This serialization step restores document-level coherence prior to prompt construction, thereby preserving the local procedural flow intended by the skill author. Finally, the textual contents of these sorted nodes are concatenated to form the definitive context for downstream LLM execution.

\paragraph{Computational Cost.}
At inference time, Stage~2 requires a single embedding call for the query, as all node embeddings are pre-computed and persistently cached during the offline parsing stage (Stage~1). The subsequent optimization process is purely numerical, evaluating cosine similarities over the cached embedding space.  The resulting cost depends on the candidate pool size and the selection budget, but requires no additional LLM calls during retrieval. In our implementation, the online retrieval overhead is dominated by the single query embedding call, which averages approximately 498 tokens, whereas the execution of the MMR loop incurs negligible latency.

\subsection{Design Rationale}
\label{sec:design-rationale}

The final architecture of \textsc{SkillPager} is guided by three empirical findings. On the full benchmark, global MMR retrieval over the node set outperforms topology-dependent traversal. In a smaller technical validation, query-time graph augmentation does not improve the efficiency--quality trade-off. On the ground-truth subset (195 skills, 975 queries), adaptive inclusion of concept nodes outperforms static alternatives.

\paragraph{From Graph Traversal to Global Retrieval.}
An earlier design used the parsed node graph directly at query time. On the full benchmark, greedy graph traversal reaches 66.73\% LLM-judged context sufficiency, while a top-$k$ embedding baseline with dependency completion reaches 57.85\%. These results suggest that neither local graph exploration nor global access without redundancy control is sufficient in this setting. By contrast, MMR-based retrieval operates over the node set, combines global access with explicit redundancy penalization, and reaches 78.89\%, marking an improvement of 12.16\% over greedy traversal. This comparison indicates that, when the parsed node graph is sparse, the main bottleneck is global selection quality rather than local traversal.

\paragraph{Graph Augmentation at Query Time.}
We further evaluate whether localized graph expansion yields incremental utility once retrieval has already been performed globally. On a validation subset comprising 19 skills and 86 ground-truth queries, we augment the MMR-selected node set by incorporating their one-hop dependency predecessors at query time. As demonstrated in Table~\ref{tab:graph-augmentation}, this modification degrades the LLM-judged context sufficiency from 69.8\% to 67.4\%, while bloating the average number of selected nodes from 17.9 to 28.1. Token reduction relative to full-document prompting drops from 62.4\% to 48.0\%. Although the overlapping 95\% confidence intervals indicate that this marginal decrease in sufficiency is statistically non-significant, the substantial inflation in context size highlights a clear efficiency penalty. We therefore retain pure global MMR selection in the final system.

\begin{table}[t]
\centering
\small
\caption{Pilot technical validation of query-time graph augmentation on top of global MMR retrieval. Results are reported on 19 skills and 86 ground-truth queries.}
\label{tab:graph-augmentation}
\begin{tabular}{lcccc}
\toprule
\textbf{Method} & \textbf{Suff. (\%)} & \textbf{95\% CI} & \textbf{Avg. Nodes} & \textbf{Token Red. (\%)} \\
\midrule
MMR & 69.8 & [59.3, 79.1] & 17.9 & 62.4 \\
MMR + 1-hop augmentation & 67.4 & [57.0, 76.7] & 28.1 & 48.0 \\
Full-document prompting & 82.6 & [74.4, 90.7] & --- & --- \\
\bottomrule
\end{tabular}
\end{table}

\paragraph{Adaptive Concept Inclusion.}
Concept nodes remain in the candidate pool during MMR selection, meaning they are evaluated dynamically for each query under the exact same MMR objective as executable nodes. We refer to this mechanism as \emph{adaptive concept inclusion}. On the ground-truth subset, restricting retrieval exclusively to executable nodes eliminates essential explanatory context and reduces context sufficiency from 86.67\% to 78.26\%, yielding an absolute performance drop of 8.41\%. Furthermore, static density-based heuristic rules that pre-commit to fixed inclusion or exclusion boundaries consistently underperform this adaptive strategy. This pivotal outcome underscores that explanatory content provides maximum utility only when it is exposed to the global retrieval objective, allowing it to be selected selectively when strictly warranted by the query. A comprehensive ablation analysis regarding concept node configurations is provided in~\S\ref{sec:concept-ablation}.

\section{Experimental Setup}
\label{sec:experiments-setup}

\subsection{Dataset}
\label{sec:dataset}

We build our benchmark from the \textsc{SkillRouter-Eval-Core} dataset, using its original \emph{easy} partition. The \emph{easy} label is part of the benchmark's own difficulty annotation rather than a partition defined for our method. The dataset provides skill records with fields including \texttt{skill\_id}, \texttt{name}, \texttt{description}, \texttt{body}, and \texttt{source}, but does not include repository-level metadata such as source URLs. We retain only records with non-empty \texttt{body} fields and apply no additional quality filtering, in order to preserve the original distribution of the benchmark.

The final benchmark contains 395 skills drawn from two sources. The first source is the ground-truth subset: among 196 records with the \texttt{gt/} prefix, we exclude one empty file (\texttt{gt/licenses}) and retain the remaining 195 skills. The second source is a pool subset sampled from 78{,}165 \texttt{pool} records. From this pool, we draw 200 skills by stratified sampling over seven namespaces (\textit{other}, \textit{development}, \textit{data}, \textit{design}, \textit{devops}, \textit{testing}, and \textit{documents}) using a fixed random seed of 42. Table~\ref{tab:dataset-composition} summarizes the benchmark composition.

\begin{table}[t]
  \centering
  \small
  \caption{Benchmark composition derived from the \textsc{SkillRouter-Eval-Core} \emph{easy} partition.}
  \label{tab:dataset-composition}
  \begin{tabular}{lrr}
    \toprule
    \textbf{Component} & \textbf{Count} & \textbf{Notes} \\
    \midrule
    Ground-truth skills & 195 & From 196 \texttt{gt/} records, excluding \texttt{gt/licenses} \\
    Pool skills         & 200 & Stratified sample from 78{,}165 \texttt{pool} records \\
    Total skills        & 395 & Records with non-empty \texttt{body} only \\
    Queries per skill   & 5   & LLM-generated \\
    Total queries       & 1{,}975 & 395 $\times$ 5 \\
    \bottomrule
  \end{tabular}
\end{table}

The resulting dataset spans diverse domains including software development, data analysis, DevOps, design, testing, and documentation. Each skill is a Markdown document ranging from 166 to 93{,}951 characters (median 4{,}163). After Stage~1 parsing, these documents yield 22{,}774 semantic nodes in total, with an average of 57.7 nodes per skill (minimum 1, maximum 529, median 38).

Table~\ref{tab:node-dist} summarizes the node-type distribution induced by Stage~1. Concept nodes account for 49.9\% of all nodes, followed by \textit{param} (22.7\%), \textit{step} (11.5\%), \textit{example} (11.1\%), \textit{precondition} (2.5\%), and \textit{error\_handling} (2.2\%). This distribution highlights the heterogeneous internal structure of skill documents and motivates the later analysis of concept-node handling in~\S\ref{sec:concept-ablation}.

\begin{table}[t]
  \centering
  \small
  \caption{Node-type distribution induced by Stage~1 over the 395-skill benchmark (22{,}774 total nodes).}
  \label{tab:node-dist}
  \begin{tabular}{lrr}
    \toprule
    \textbf{Type} & \textbf{Count} & \textbf{Proportion (\%)} \\
    \midrule
    concept         & 11{,}369 & 49.9 \\
    param           & 5{,}166  & 22.7 \\
    step            & 2{,}617  & 11.5 \\
    example         & 2{,}530  & 11.1 \\
    precondition    & 580      & 2.5 \\
    error\_handling & 512      & 2.2 \\
    \midrule
    total           & 22{,}774 & 100.0 \\
    \bottomrule
  \end{tabular}
\end{table}

For evaluation, we generate five task-oriented queries per skill entirely with an LLM, yielding 1{,}975 queries in total. Query generation uses DeepSeek-V4-Flash, with each skill's \texttt{name} and full \texttt{body} provided as input. The prompt requires exactly five queries per skill, each phrased as a concrete agent request of one to three sentences, strictly grounded in the scope of the skill document, and collectively covering different usage angles and difficulty levels. Generation is executed with 8-way parallelism and up to three retries with exponential backoff on failure, and the resulting queries are stored as one JSON file per skill. Because these queries are LLM-generated rather than human-authored, this benchmark should be interpreted as measuring controlled query-conditioned retrieval behavior rather than fully natural user traffic.

\subsection{Evaluation Protocol}
\label{sec:eval-protocol}

We evaluate each method by \emph{context sufficiency}. Given a query and a selected context, the judge determines whether the selected fragments contain the critical information needed to complete the task within the scope of the skill. This metric evaluates retrieval sufficiency, not end-to-end agent success.

The judge applies a strict binary criterion: \emph{sufficient} if the selected context contains the required operational steps, parameters, and essential constraints for the query, and \emph{insufficient} otherwise. Each judgment is repeated three times with deterministic decoding, and the final label is given by majority vote.

We validate this protocol against human annotation on a stratified sample of 96 query--context pairs. The sample is balanced across four methods, two judge labels, and two vote patterns (3-0 vs.\ 2-1), with 6 instances per cell. The majority-vote judge reaches 82.29\% agreement with the human annotations, with Cohen's $\kappa = 0.646$. Agreement is higher on 3-0 cases (85.42\%, $\kappa = 0.708$) than on 2-1 cases (79.17\%, $\kappa = 0.583$).

We therefore use LLM-judged context sufficiency as a human-validated proxy for comparative evaluation under this protocol. The confidence intervals reported in ~\S\ref{sec:results} quantify uncertainty under this judge protocol.

\subsection{Baselines}
\label{sec:baselines}

We compare \textsc{SkillPager} against four baselines. Three node-based baselines operate on the same Stage~1 representation, using the same parsed node set, node embeddings (BAAI/bge-m3), and cached graph structure, and differ only in how nodes are selected at query time. The chunk baseline replaces typed semantic nodes with fixed-length chunks to isolate granularity effects.

\begin{itemize}
  \item \textbf{Full-document prompting}: All parsed nodes are included in the context. We treat this setting as a full-context reference condition rather than an oracle upper bound.
  
  \item \textbf{Greedy graph traversal}: Starting from an entry node, the retriever greedily follows graph edges to collect additional nodes until a gain threshold is met. This baseline represents topology-guided local traversal on the inferred skill graph.
  
  \item \textbf{Top-$k$ + dependency completion}: The top-$k$ nodes by query--node cosine similarity are first selected, after which graph predecessors of the selected nodes are added as dependency completion. This baseline tests whether global embedding access alone is sufficient when combined with local graph augmentation.
  
  \item \textbf{Naive chunk + MMR}: We partition each skill document into fixed 256-token chunks and apply the same MMR-based selection procedure used by \textsc{SkillPager}. This baseline controls for the retrieval objective and tests whether the gain comes from typed semantic nodes or simply from global MMR retrieval over arbitrary text segments.
  
\end{itemize}

These baselines isolate three factors in the design space: graph topology, retrieval objective, and retrieval granularity. Greedy traversal and top-$k$ + dependency completion operate on the same Stage~1 node representation as \textsc{SkillPager} and test the effect of topology-dependent retrieval. Naive chunk + MMR keeps the global MMR objective but replaces typed semantic nodes with fixed-length chunks, isolating the effect of retrieval granularity.

\subsection{Implementation Details}
\label{sec:implementation}

We use BAAI/bge-m3~\citep{chen2024bge} to compute 1{,}024-dimensional embeddings for both semantic nodes (offline, cached) and queries (online, per request). The benchmark is predominantly English, but it is not strictly English-only: 383 of 395 skills are English-dominant, 11 contain mixed-language content, and 1 is non-English-dominant. We therefore use a single embedding model that covers the full benchmark without introducing language-specific retrieval branches. All similarity computations use cosine similarity. Node classification in Stage~1 is performed with DeepSeek-V4-Flash, using a batch size of 20.

Each automated judgment is repeated three times via DeepSeek-V4-Pro and aggregated through majority voting, adhering to the protocol established in~\S\ref{sec:eval-protocol}. In Stage~2, we set the MMR parameter to $\lambda=0.7$ and use the two-tier dynamic budget described in~\S\ref{sec:d2}: $k=20$ for skills containing at most 100 nodes and $k=60$ otherwise.

\section{Results and Analysis}
\label{sec:results}

We organize our evaluation around three research questions that examine the system from semantic parsing quality through retrieval effectiveness to hyperparameter sensitivity.

\subsection{RQ1: Does typed parsing produce a sufficiently reliable retrieval representation?}
\label{sec:rq1}

We first examine whether Stage~1 produces a sufficiently reliable typed representation for downstream retrieval. If the six-class taxonomy (\texttt{concept}, \texttt{step}, \texttt{param}, \texttt{example}, \texttt{precondition}, \texttt{error\_handling}) is substantially noisy, then Stage~2 operates over a corrupted search space. We evaluate Stage~1 through three complementary lenses: a small-scale human evaluation, a qualitative parsed-graph example, and a downstream concept ablation.

\subsubsection{Direct Evidence: Human Evaluation}

To provide direct evidence, we conduct a small-scale human evaluation on a stratified sample of 96 parsed nodes, balanced across predicted node types, skill sources, and skill sizes. Against human labels, the parser achieves 79.17\% accuracy and 79.27\% macro-F1. The strongest categories are \texttt{precondition} and \texttt{error\_handling}, while the main confusions occur between semantically adjacent categories, especially \texttt{concept}, \texttt{param}, and \texttt{example}. These results indicate that the parser is sufficiently accurate to support typed retrieval, while still leaving room for improvement in intrinsic label fidelity.

\subsubsection{Qualitative Parse Example}

\begin{figure}[t]
  \centering
  \includegraphics[width=\linewidth]{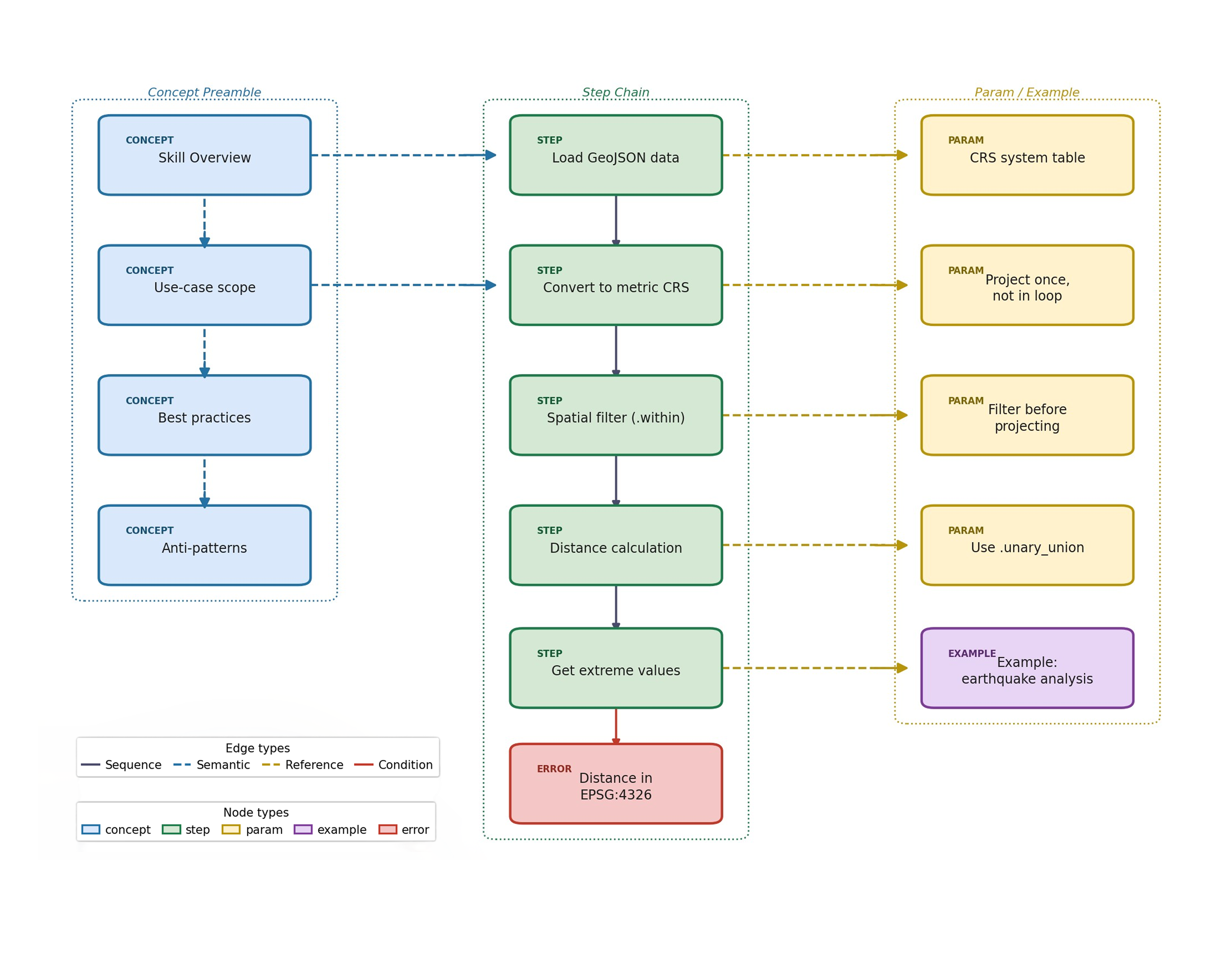}
  \caption{Representative parsed node graph of a 42-node skill. Colors indicate node types: blue = concept, green = step, orange = param, and purple = example. Edges represent structural relationships inferred during Stage~1.}
  \label{fig:dag-example}
\end{figure}

Figure~\ref{fig:dag-example} shows a representative parsed node graph from a medium-complexity skill containing 42 nodes. The parser separates the document into a concept-oriented preamble, followed by executable steps and their associated supporting content. Parameter nodes are attached to the steps that consume them, and error-handling nodes appear near the procedures they qualify. This example illustrates the structural relations inferred during Stage~1. While query-time retrieval in \textsc{SkillPager} operates over the vectorized flat node set to maximize online efficiency, this underlying topology remains crucial as it defines the structural search space boundaries and provides the backbone for topology-dependent baselines.

\subsubsection{Downstream Utility: Concept Ablation}

We next assess the utility of the typed representation through concept ablation (Table~\ref{tab:concept-ablation}; see also~\S\ref{sec:concept-ablation}). In the \textbf{exec\_only} variant, all nodes labeled as \textit{concept} are removed before retrieval, leaving only the five executable types. On the ground-truth subset, \textbf{exec\_only} reaches 78.26\% LLM-judged context sufficiency, compared with 86.67\% for the \textbf{adaptive} variant that retains concept nodes in the candidate pool. This result shows that the parser exposes explanatory content that is often useful for query resolution, even though such content is not itself executable. It should therefore be interpreted as evidence of downstream utility rather than as a direct measure of intrinsic parsing accuracy.

Taken together, the human evaluation, the qualitative parsed-graph example, and the concept ablation support the conclusion that Stage~1 produces a useful and sufficiently reliable retrieval representation, though not a near-perfect semantic parser in the benchmark sense.
\subsection{RQ2: How Effective Is the Navigation Strategy?}
\label{sec:rq2}

\subsubsection{Main Experiment: Baseline Comparison}
\label{sec:rq2-main}

As shown in Table~\ref{tab:node-dist}, concept nodes account for 49.9\% of all parsed nodes. Therefore full-document prompting pays a substantial cost for explanatory content on every query. Table~\ref{tab:main-results} tests whether query-adaptive navigation yields a better efficiency and sufficiency trade-off. \textsc{SkillPager} achieves 78.89\% context sufficiency, compared with 82.23\% for full-document prompting, while reducing prompt tokens by 47.04\%. Crucially, the bootstrap 95\% confidence intervals for these two front-runners exhibit a tight, marginal overlap strictly confined to the [80.51\%, 80.61\%] interval. This statistical proximity demonstrates that \textsc{SkillPager} successfully establishes statistical non-inferiority against the exhaustive full-context reference condition, preserving almost all execution-critical directives at a near-halved inference budget.

\begin{table}[H]
\centering
\small
\setlength{\tabcolsep}{5pt}
\caption{Main results on the full benchmark (395 skills, 1{,}975 queries). \emph{Suff.} denotes LLM-judged context sufficiency (3-vote majority). \emph{Token Red.} denotes aggregate fragment token reduction relative to full-document prompting. Bootstrap 95\% confidence intervals are reported for all methods.}
\label{tab:main-results}
\begin{tabular}{p{4.2cm}cccc}
\toprule
\textbf{Method} & \textbf{Suff. (\%)} & \textbf{95\% CI} & \textbf{Avg. Nodes} & \textbf{Token Red. (\%)} \\
\midrule
Full-document prompt & 82.23 & [80.51, 83.90] & 57.43 & --- \\
\midrule
Top-$k$ + dependency & 57.85 & [55.65, 59.95] & 14.80 & 36.00 \\
Greedy graph & 66.73 & [64.66, 68.76] & 21.59 & 45.10 \\
Naive chunk + MMR & 81.77 & [80.05, 83.44] & 7.09 & $-$28.81 \\
\midrule
\textbf{\textsc{SkillPager} (Ours)} & \textbf{78.89} & \textbf{[77.01, 80.61]} & \textbf{24.68} & \textbf{47.04} \\
\bottomrule
\end{tabular}
\end{table}

\paragraph{Greedy graph.}
Greedy graph selects 21.59 nodes on average, only 3.09 fewer than \textsc{SkillPager}, yet its context sufficiency is 12.16\% lower. This gap is therefore not explained by node count alone. Because greedy traversal expands locally from a single entry node, it is vulnerable to entry-point bias when relevant information is distributed across distant regions of a skill. The comparison suggests that the main advantage of \textsc{SkillPager} lies in global selection quality rather than larger retrieved context.

\paragraph{Top-$k$ + dependency completion.}
Top-$k$ + dependency completion reaches 57.85\% context sufficiency.
Although it has global embedding access, it does not explicitly control redundancy during selection. As a result, semantically overlapping nodes can consume budget while other relevant aspects of the query remain uncovered. This comparison suggests that global access alone is insufficient without explicit diversity control.

\paragraph{Naive chunk + MMR.}
Naive chunk + MMR reaches 81.77\% sufficiency, matching the full-document reference, yet increases total injected tokens by 28.81\%. This engineering overhead is primarily driven by granularity mismatch: fixed 256-token linear chunk windows are inherently coarser than our granular semantic nodes, meaning each retrieved chunk forces the injection of substantial trailing irrelevant content. \textsc{SkillPager} achieves 47.04\% token reduction with the same retrieval algorithm by operating on semantically typed nodes, demonstrating that token efficiency depends on retrieval unit granularity, not the retrieval algorithm itself.

\subsubsection{Concept Candidate Pool Design}
\label{sec:concept-ablation}

Table~\ref{tab:concept-ablation} isolates the contribution of concept nodes on the ground-truth subset. This subset comprises 195 skills from the \textsc{SkillRouter-Eval-Core} \emph{easy} partition, where the \emph{easy} label comes from the benchmark's own difficulty annotation rather than any selection made for our method. Each skill in this subset has at least one curated reference answer, which enables controlled ablation and sensitivity analysis. The subset was fixed before any hyperparameter or design decisions were made. The remaining 200 pool skills were excluded from parameter selection and are used only in the main benchmark (Table~\ref{tab:main-results}), which reduces the risk of post-hoc tuning on the ablation population.

\begin{table}[H]
\centering
\caption{Concept node ablation on the ground-truth subset. \emph{Adaptive} keeps concept nodes in the MMR candidate pool without pre-filtering; \emph{density adaptive} excludes concept nodes from skills whose concept ratio exceeds a threshold $\tau$. Bootstrap 95\% confidence intervals are shown in brackets. Judge vote strong-agreement rate (3--0 votes): 89.9\%; borderline rate (2--1 votes): 10.1\%.}
\label{tab:concept-ablation}
\begin{tabular}{lccc}
\toprule
\textbf{Variant} & \textbf{Suff. (\%)} & \textbf{95\% CI} & \textbf{Avg. Nodes} \\
\midrule
Concept only & 20.21 & [17.74, 22.77] & 13.85 \\
Exec only & 78.26 & [75.59, 80.82] & 15.74 \\
\midrule
Density adaptive ($\tau=0.3$) & 63.38 & [60.31, 66.46] & 22.05 \\
Density adaptive ($\tau=0.4$) & 62.56 & [59.59, 65.64] & 21.66 \\
Density adaptive ($\tau=0.5$) & 62.46 & [59.49, 65.54] & 20.07 \\
\midrule
\textbf{Adaptive (ours)} & \textbf{86.67} & \textbf{[84.51, 88.72]} & \textbf{22.34} \\
Full-document prompting & 88.41 & [86.36, 90.36] & 47.87 \\
\bottomrule
\end{tabular}
\end{table}

\paragraph{Concept nodes are insufficient alone but useful in combination.}
Concept-only retrieval reaches only 20.21\%, confirming that explanatory content cannot substitute for executable instructions. However, keeping concept nodes in the candidate pool improves performance substantially: \textbf{adaptive} reaches 86.67\%, exceeding \textbf{exec\_only} by 8.41\%. To quantify when concept nodes matter, we identify the subset of queries for which \textbf{adaptive} succeeds but \textbf{exec\_only} fails. This set contains 113 of 975 queries, and within these queries the selected context contains 54\% concept nodes on average. The gain is therefore concentrated in a specific slice of the benchmark that depends on terminology, architectural context, or design rationale.

\paragraph{Density-based gating underperforms consistently.}
All three density-adaptive variants ($\tau \in \{0.3, 0.4, 0.5\}$) fall between 62.46\% and 63.38\%, substantially below even \textbf{exec\_only}. The reason is that density gating removes concept nodes before relevance is evaluated. For skills above the threshold, the retriever must select from an impoverished candidate pool whose remaining executable nodes may not span the semantic space required by the query. In this setting, corpus-level density statistics are a poor proxy for per-query usefulness.

\paragraph{Design implication.}
The adaptive strategy works because it defers the inclusion decision to the retrieval objective itself. Under MMR, concept nodes are selected only when they contribute query-relevant and non-redundant information; otherwise, they are naturally excluded. This suggests a broader design principle for typed intra-document retrieval: potentially useful supporting content should remain available to the scorer, rather than being removed by static heuristics before retrieval begins.

\begin{figure}[H]
  \centering
  \includegraphics[width=\linewidth]{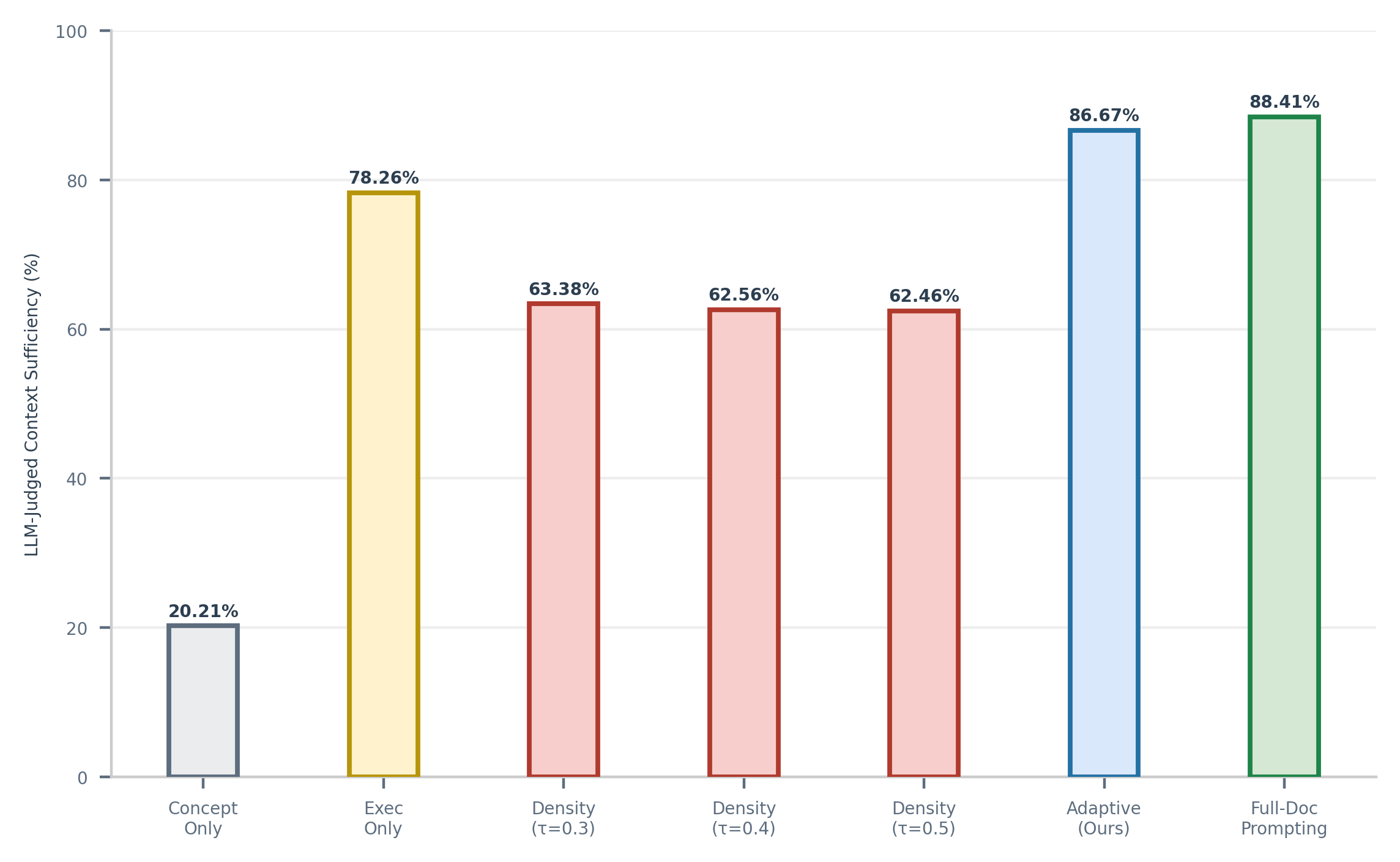}
  \caption{Concept ablation results. The adaptive strategy approaches the full-document reference while using far fewer nodes, whereas density-based pre-filtering performs worse than even the executable-only baseline.}
  \label{fig:concept-ablation}
\end{figure}
\subsection{RQ3: How Do Hyperparameters Behave and What Motivates the Dynamic Budget?}
\label{sec:rq3}

All hyperparameter sweeps ($\lambda$ and $k$) are conducted on the ground-truth subset. The remaining 200 pool skills are excluded from parameter selection and appear only in the main benchmark results. This separation reduces the risk that the operating points reported below reflect post-hoc tuning on the primary evaluation population.

\subsubsection{Relevance--Diversity Trade-off ($\lambda$)}
\label{sec:lambda-sweep}

Table~\ref{tab:lambda-sweep} reports the effect of $\lambda$ on the ground-truth subset. Larger $\lambda$ places more weight on query relevance relative to the diversity penalty in the MMR objective.

\begin{table}[H]
\centering
\caption{Effect of $\lambda$ on LLM-judged context sufficiency (ground-truth subset). Bootstrap 95\% confidence intervals are computed over queries ($B=10{,}000$, seed 42).}
\label{tab:lambda-sweep}
\begin{tabular}{lccc}
\toprule
$\lambda$ & \textbf{Suff. (\%)} & \textbf{95\% CI} & \textbf{Avg. Nodes} \\
\midrule
0.3 & 42.56 & [39.49, 45.74] & $\sim$1.0 \\
0.5 & 54.05 & [50.77, 57.13] & 3.97 \\
\textbf{0.7} & \textbf{85.95} & \textbf{[83.79, 88.10]} & \textbf{22.34} \\
0.9 & 86.05 & [83.90, 88.21] & 22.35 \\
\bottomrule
\end{tabular}
\end{table}

The transition from $\lambda=0.5$ to $\lambda=0.7$ produces a 31.90\% jump, from 54.05\% to 85.95\%, while $\lambda=0.7 \to 0.9$ changes performance by only 0.10\%, with heavily overlapping confidence intervals (0.7: [83.79, 88.10]; 0.9: [83.90, 88.21]). This pattern reflects the joint behavior of the MMR relevance-diversity weighting and the early stopping criterion (terminate when all candidate scores fall below zero). At $\lambda=0.3$, the diversity penalty is so dominant that the first selected node immediately drives all remaining candidate scores below zero; the algorithm terminates after a mean of $\sim$1.0 node, effectively retrieving a single entry regardless of query complexity and collapsing context sufficiency to 42.56\%. At $\lambda=0.5$, selection terminates after only 3.97 nodes on average, indicating that the diversity penalty is too strong to support budget utilization. At $\lambda=0.7$, the retriever enters a stable regime with substantially higher coverage. Further increasing $\lambda$ provides no clear additional gain.

\begin{figure}[t]
  \centering
  \includegraphics[width=0.9\linewidth]{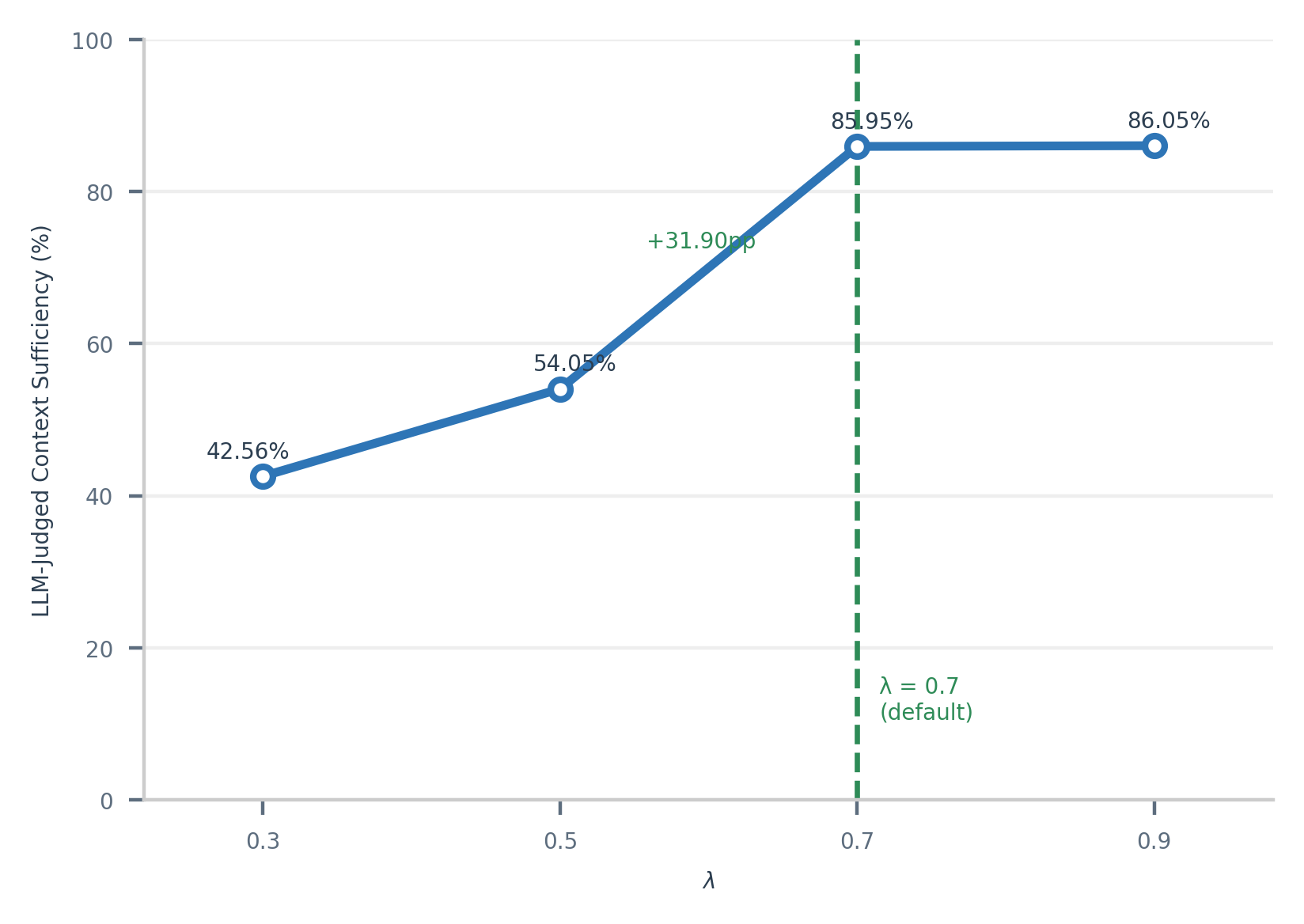}
  \caption{LLM-judged context sufficiency as a function of $\lambda$. The sharp transition between 0.5 and 0.7 reflects the point at which the relevance term becomes strong enough to support budget utilization under the early stopping rule.}
  \label{fig:lambda-sweep}
\end{figure}

\subsubsection{Selection Budget ($k$) and Pareto Efficiency}
\label{sec:k-sweep}

\begin{figure}[t]
  \centering
  \includegraphics[width=\linewidth]{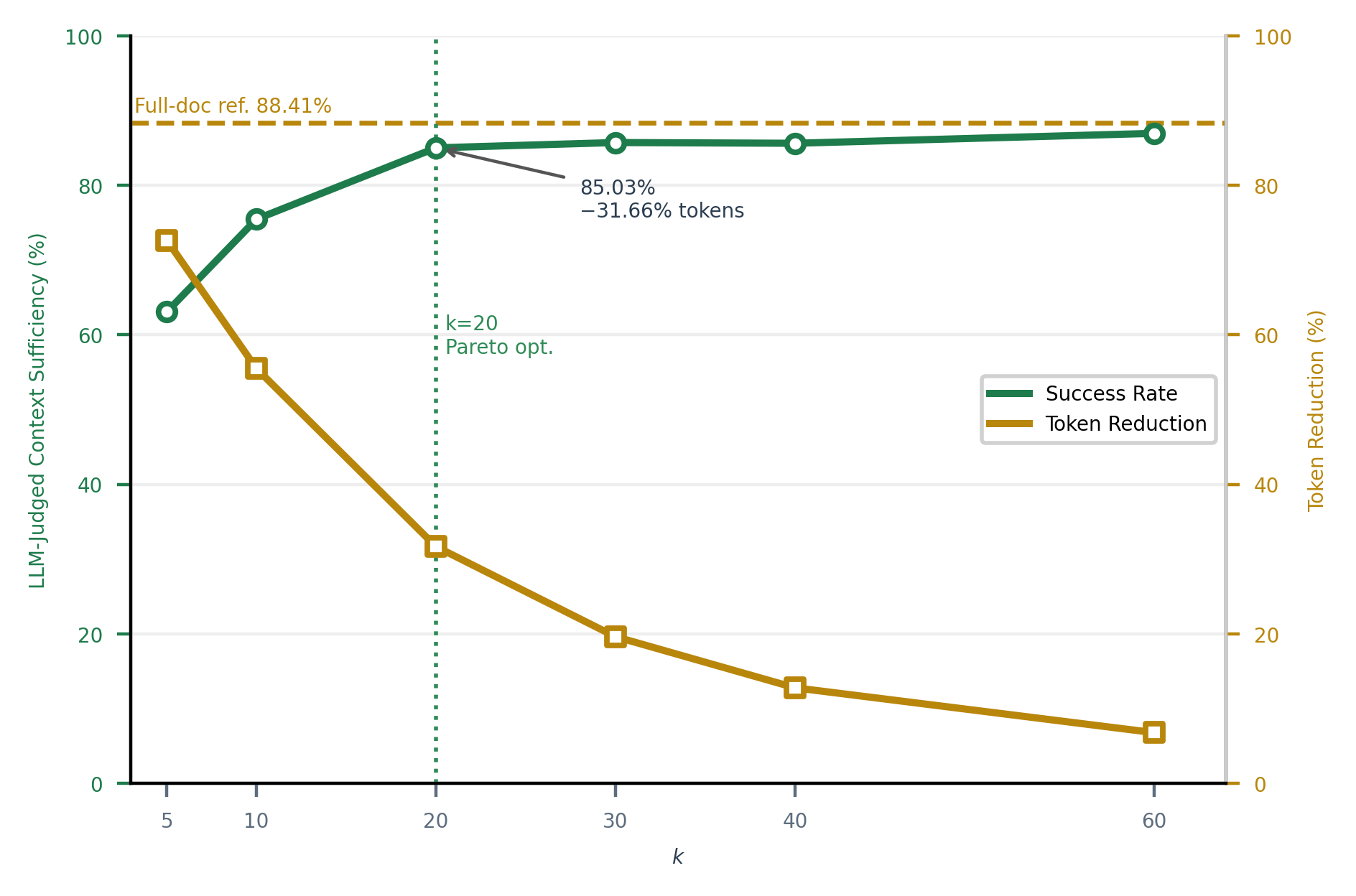}
  \caption{Trade-off between LLM-judged context sufficiency (left axis, solid) and token reduction (right axis, dashed) as a function of $k$. The marked $k=20$ operating point achieves 85.03\% sufficiency at 31.66\% token reduction, and larger budgets provide only limited additional gains.}
  \label{fig:k-sweep}
\end{figure}

Table~\ref{tab:k-sweep} and Figure~\ref{fig:k-sweep} summarize the effect of the selection budget $k$ on the ground-truth subset.

\begin{table}[t]
\centering
\caption{Effect of selection budget $k$ on the ground-truth subset. Token reduction is measured relative to full-document prompting on this subset.}
\label{tab:k-sweep}
\begin{tabular}{lcccc}
\toprule
$k$ & \textbf{Suff. (\%)} & \textbf{95\% CI} & \textbf{Avg. Nodes} & \textbf{Token Red. (\%)} \\
\midrule
5 & 63.08 & {[}60.00, 66.05{]} & 4.99 & 72.67 \\
10 & 75.49 & {[}72.72, 78.15{]} & 9.70 & 55.48 \\
\textbf{20} & \textbf{85.03} & \textbf{{[}82.77, 87.28{]}} & \textbf{18.24} & \textbf{31.66} \\
30 & 85.74 & {[}83.49, 87.90{]} & 24.57 & 19.61 \\
40 & 85.64 & {[}83.38, 87.79{]} & 29.27 & 12.77 \\
60 & 86.97 & {[}84.82, 89.03{]} & 35.08 & 6.77 \\
Full & 88.41 & {[}86.36, 90.36{]} & 47.87 & — \\
\bottomrule
\end{tabular}
\\[4pt]
\end{table}

The curve exhibits clear diminishing returns beyond $k=20$: increasing the budget from $k=20$ to $k=30$ improves LLM-judged context sufficiency by only 0.71\% while reducing token savings by 12.05\%. This makes $k=20$ a practical Pareto-efficient operating point for most skills in our benchmark. 

\subsubsection{Dynamic Budget Strategy}
\label{sec:rq3-dynamic}

The fixed $k=20$ operating point is effective on average, but the benchmark is heterogeneous in skill size: 84.10\% of skills contain at most 100 nodes (median 38), whereas the remaining 15.90\% contain more than 100 nodes, with a maximum of 529. For smaller skills, $k=20$ already covers a substantial fraction of the node set and is typically sufficient. For larger skills, the same budget may cover less than one fifth of the node set, increasing the risk of under-retrieval.

We therefore adopt a two-tier dynamic budget:
\begin{itemize}
  \item skills with $|\mathcal{V}(D)| \leq 100$: $k=20$;
  \item skills with $|\mathcal{V}(D)| > 100$: $k=60$.
\end{itemize}

A size-stratified $k$-sweep is consistent with this choice. For skills with at most 100 nodes, increasing the budget from $k=20$ to $k=60$ yields only a modest gain, from 86.40\% to 87.89\%. For skills with more than 100 nodes, the same change raises sufficiency from 73.00\% to 79.00\%. We therefore use $k=60$ for large skills as a conservative choice to reduce the risk of under-retrieval, while noting that the aggregate gain beyond $k=20$ remains modest. The 100-node boundary lies near the upper tail of the skill-size distribution (84th percentile), and nearby thresholds such as 80 and 120 produce similar behavior, changing aggregate sufficiency by less than 0.5\% on the full benchmark. The two-tier budget is therefore stable over a reasonable range of threshold choices.

\section{Conclusion}
\label{sec:conclusion}

This paper studies intra-skill retrieval for skill-based LLM agents. We introduce \textsc{SkillPager}, a two-stage framework that parses a skill document into typed semantic nodes and then performs global query-adaptive retrieval over the resulting node set with an MMR objective. Across 395 skills and 1{,}975 queries, \textsc{SkillPager} achieves 78.89\% LLM-judged context sufficiency while reducing prompt tokens by 47.04\% relative to full-document prompting. A granularity ablation further shows that applying the same MMR retrieval algorithm to raw chunks reaches comparable sufficiency but increases total token cost, indicating that the main efficiency gain comes from the typed node representation rather than from the retrieval objective alone. Among graph-based baselines, \textsc{SkillPager} improves over the strongest baseline by 12.16\%. The main empirical lesson is that useful skill context is compositional rather than document-level: supporting content should remain available to the retriever and be included only when warranted by the query.

The study also has important limitations. The current evaluation targets LLM-judged context sufficiency rather than downstream execution success. We partially mitigate this risk through the human validation in~\S\ref{sec:eval-protocol}: on 96 stratified query--context pairs, the majority-vote judge reaches 82.29\% agreement with human annotation ($\kappa = 0.646$), supporting its use as a comparative proxy under our protocol. However, the evaluation still uses only the \emph{easy} partition of SkillRouter-Eval-Core and relies on synthetic queries generated by DeepSeek-V4-Flash under a single embedding family. This validation inherently addresses the reliability of the automated judge rather than the broader query distribution. While synthetic generation provides a scalable and grounded testbed for closed-loop evaluation, it introduces a potential lexical positioning bias where synthetic phrases might systematically align closer with the embedding space than idiosyncratic human search noise. Consequently, our findings should be interpreted as a rigorous benchmark under standard agent-traffic configurations, while leaving out-of-distribution human intent alignment as an active future direction. The current setup also assumes that the relevant skill document is already known, and therefore does not cover routing errors or multi-skill settings where context must be composed across several partially relevant skills. Lastly, \textsc{SkillPager} optimizes online steady-state execution efficiency by shifting the structural parsing workload offline. This introduces a one-time offline preprocessing overhead during initial skill onboarding, rendering the framework exceptionally well-suited for stable production repositories where partitioned semantic assets are frequently reused post-parsing.

These limitations point to several next steps: evaluation on harder benchmark regimes, human-authored or execution-derived queries, end-to-end agent benchmarks, and joint routing-plus-retrieval settings; lighter-weight Stage~1 parsing for lower onboarding cost; and stronger chunk-level or hybrid baselines. More broadly, we view intra-skill retrieval as an increasingly important interface problem for agent systems as skill libraries continue to scale.

\bibliography{main}
\bibliographystyle{rlc}

\end{document}